\documentclass[aip, amsmath,amssymb, reprint]{revtex4-2}
\usepackage{natbib}
\usepackage[hyperfootnotes=true]{hyperref} %Testing
\usepackage{siunitx} % allows for unit quantities
\usepackage{graphicx}

\begin{document}

\renewcommand{\arraystretch}{2}

\title{Applying Machine Learning Methods to Laser Acceleration of Protons: \\Synthetic Data for Exploring the High Repetition Rate Regime }

%  \orcid{0009-0002-8877-1864}  \orcid{0000-0003-1648-580X}
\author{John J. Felice}
\affiliation{Department of Physics, The Ohio State University, Columbus, OH, USA}
\author{Ronak Desai}
\email{desai.458@osu.edu}
\affiliation{Department of Physics, The Ohio State University, Columbus, OH, USA}
\author{Nathaniel Tamminga}
\affiliation{Department of Physics, The Ohio State University, Columbus, OH, USA}
\author{Joseph R. Smith}
\affiliation{Department of Physics, Marietta College, Marietta, OH, USA}
\author{Alona Kryshchenko}
\affiliation{Department of Mathematics, California State University Channel Islands, Camarillo, CA, USA}
\author{Chris Orban}
\affiliation{Department of Physics, The Ohio State University, Columbus, OH, USA}
\author{Michael L. Dexter}
\affiliation{Department of Engineering Physics, Air Force Institute of Technology, Wright-Patterson AFB, OH, USA}
\author{Anil K. Patnaik}
\affiliation{Department of Engineering Physics, Air Force Institute of Technology, Wright-Patterson AFB, OH, USA}

% Abstract
\begin{abstract}
Advances in ultra-intense laser technology have increased repetition rates and average power for chirped-pulse laser systems, which offers a promising solution for many applications including energetic proton sources. An important challenge is the need to optimize and control the proton source by varying some of the many degrees of freedom inherent to the laser-plasma interactions. Machine learning can play an important role in this task, as our work examines. Building on our earlier work in Desai et al. 2024, we generate a large $\sim$1.5 million data point synthetic data set for proton acceleration using a physics-informed analytic model that we improved to include pre-pulse physics. Then, we train different machine learning methods on this data set to determine which methods perform efficiently and accurately. Generally, we find that quasi-real-time training of neural network models using single shot data from a kHz repetition rate ultra-intense laser system should typically be feasible on a single GPU. We also find that a less sophisticated model like a polynomial regression can be trained even faster and that the accuracy of these models is still good enough to be useful. We provide our source code and example synthetic data for others to test new machine learning methods and approaches to automated learning in this regime.
\end{abstract}

% Keywords
% \keywords{Machine learning; laser-driven ion acceleration}

\maketitle

\section{Introduction}
  \label{sec:intro}

As ultra-intense laser systems improve and research groups translate proof-of-concept experiments into applications, the repetition rate of these laser systems becomes an important frontier. Survey papers anticipate growth in this area and outline challenges and opportunities inherent in the high repetition rate regime \citep{Hooker2013,Heuer_etal2022,Feister_etal2023}. One area that will benefit from improvements in repetition rate is the laser acceleration of protons and ions, for example through Target Normal Sheath Acceleration (TNSA) \cite{Clark_etal2000,Hatchett_etal2000,Snavely_etal2000,Passoni_etal2010} and other mechanisms \cite{Badziak_2018}.

The problem of optimizing and controlling proton acceleration on ultra-intense laser systems is something that multiple groups are working to address. \citet{Loughran_etal2023} describe efforts to control and optimize proton acceleration on a 1~Hz repetition rate laser system using Bayesian optimization. \citet{Ma_etal2021} describes a collaboration between researchers at Lawrence Livermore National Laboratory and Colorado State University to use machine learning methods to optimize and control proton acceleration with the ALEPH laser system at Colorado State University, which operates at a repetition rate of 1-3 Hz. The Extreme Light Infrastructure (ELI) facility in Bucharest can operate at up to 10~Hz repetition rate \cite{ELI}  and other systems with lower intensity and energy are already operating at a kHz and performing proton acceleration experiments (e.g. \citet{Morrison_etal2018}).

Specific challenges addressed in this paper are processing, in quasi-real time, the large volume of proton acceleration data that a kHz repetition rate experiment would produce and training a machine learning (ML) model to perform optimization and control tasks using those data. This processing should require no more than modest computer equipment, like a desktop computer with a single GPU. Training ML models with desktop-class hardware has become increasingly feasible as consumer technology has advanced, and restricting hardware for training to locally accessible machines (as opposed to cluster computing) allows the methods presented in this paper to be accessible in data-sensitive scenarios, such as when working with classified data that cannot be processed on the Internet. With similar goals and this same computational constraint, in an earlier paper \cite{Desai} we used a physics-motivated model for proton acceleration based on \citet{Fuchs2005} to generate 20,000 synthetic data points, then training different ML models on those data. We examined the accuracy, performance, and memory consumption of the ML models. Interestingly, we found that some ML models produced misleading results for the optimum conditions for accelerating protons. This result motivates continued work to train ML models on increasingly realistic synthetic data as will be described later in this paper.

Our earlier paper was limited in assuming no modification of the target from the ``pre-pulse" heating that occurs before the arrival of the main laser pulse, which is well known to affect the proton energy spectrum (e.g. \cite{Morrison_etal2018, Loughran_etal2023}). We also restricted our synthetic data set to 20,000 training and 5,000 testing points, even though a kHz ultra-intense laser experiment can produce millions of shots in less than an hour and store data from each of these shots \cite{Feister_etal2023}. Our synthetic data set randomly sampled the parameter space of target thickness, target position and pulse energy even though a real experiment would be designed to explore the parameter space in a very specific and sequential way. 

The work here addresses many of these limitations in order to more closely connect to experiments in this regime. We aim to demonstrate a procedure for using current widespread ML models to identify optimal conditions for proton acceleration via TNSA by generating a synthetic data set demonstrating many of the same identifying features observed in experimental data (as will be discussed further in \autoref{sec:synthetic}), containing an amount of data theoretically able to be collected by a high-repetition-rate system in under an hour.  This procedure can be readily extended to more feature-rich data obtained from experiments using the same principles demonstrated in this work. Section~\ref{sec:synthetic} describes our improved synthetic data model including details of the pre-pulse physics that were added. Section~\ref{sec:ml} describes different ML methods that were used. We also describe a relatively unsophisticated and fast-performing polynomial regression model with many fewer free parameters than the ML models that was also trained on the synthetic data. In Section~\ref{sec:results} we consider the accuracy and training time of the models. In Section~\ref{sec:opt} we use the trained models to perform example optimization tasks. In Section~\ref{sec:concl} we summarize and conclude. Appendix~\ref{ap:hyper} provides information on hyperparameter choices. Appendix~\ref{ap:campaigns} describes results from training the models with synthetic data produced with a more realistic parameter scan. Importantly, we include the source code and example data sets used in this paper to facilitate efforts to build from our approach such as exploring the performance of other ML models, or further improvements to the physics model or noise model that we use.

\section{Synthetic Data}
\label{sec:synthetic}

\subsection{Modified Fuchs et al. Model}
\label{sec:fuchs}

To test the ability of machine learning models to predict the acceleration of protons, we generate a synthetic data set that contains pairs of inputs and outputs, where the inputs describe the independent variables of laser and target parameters and the outputs describe aspects of the predicted proton energy distribution that the experimenter would measure as dependent variables. In what follows, one pair of input and output conditions is often referred to as a data point. We use an analytical model to produce data instead of simulations or experimental results because of the significantly lower time and resource costs to generate large data sets, which was crucial in producing over 1~million data points in a short time frame. For our purposes, it is sufficient that this model is based on physical arguments and replicates important features observed in experimental data. It is not necessary that the model perfectly predicts the results of proton acceleration experiments because our aim is to evaluate the predictive capacity and computational efficiency of different machine learning and statistical models that have been trained on these data. 

We generate synthetic data by modifying a physical model described by \citet{Fuchs2005} which draws from \citet{Mora_2003}, which can predict a proton energy distribution from the specifications of a laser pulse interacting with a flat target.
Besides \citet{Fuchs2005} there are a number of other semi-analytic models that could be used for this task \cite{Schreiber_etal2006,Passoni_Lontano2008,Passoni_etal2010,Zimmer_etal2021}. We chose \citet{Fuchs2005} because it is relatively simple, and because it is reasonably well-suited to predict the results of proton acceleration in the intensity range where TNSA is the relevant mechanism \cite{Clark_etal2000,Hatchett_etal2000,Snavely_etal2000,Passoni_etal2010}.

To better equip the \citet{Fuchs2005} model to accurately represent features observed in experimental data, we made a few modifications, including (1) changing the proportionality between the acceleration time and the laser pulse duration, (2) allowing the target to expand due to the pre-heating of a pre-pulse and (3) allowing the energy of the main laser pulse to decrease due to the propagation through the pre-plasma.

In \citet{Fuchs2005}, the proportionality constant between the acceleration timescale of the protons ($\tau_{\rm acc}$) and the laser pulse duration ($\tau_{\rm laser}$) was a free parameter that was set by experimental data that was available at the time. Changing this proportionality strongly affects the peak proton energy and in \citet{Desai} we adjusted it to avoid under-predicting the observed peak proton energy in a mJ-class high repetition rate experiment described by \citet{Morrison_etal2018}. In this work, we assume  
\begin{equation}
\tau_{\rm acc} = 25.0 \cdot \tau_{\rm laser}.    
\end{equation}
This assumes a significantly larger proportionality constant than was used in \citet{Desai} which did not assume any pre-pulse physics. We found that when the target is allowed to expand (as will be described in this section), the increased effective target thickness causes the model to predict lower proton energies than would be expected, so we increased the proportionality to maintain the correspondence to experimental results \cite{Morrison_etal2018}. The proportionality we chose is still within the range that \citet{Djordjević_etal2021} found when fitting to ensembles of 1D PIC simulations.

Because the \citet{Fuchs2005} model predicts both the maximum proton energy and the proton energy spectrum, it is straightforward to integrate the proton energy spectrum to calculate both the total energy in protons and the average energy in protons. Because the proton energy spectrum is well behaved, rather than introducing a low energy cutoff, we calculated these quantities by integrating to zero proton kinetic energy. Our modified \citet{Fuchs2005} model therefore predicts the maximum proton kinetic energy, total energy in protons, and average energy in protons for a given set of input conditions. 

\begin{figure}[t]
    \centering
    \includegraphics[width=3in]{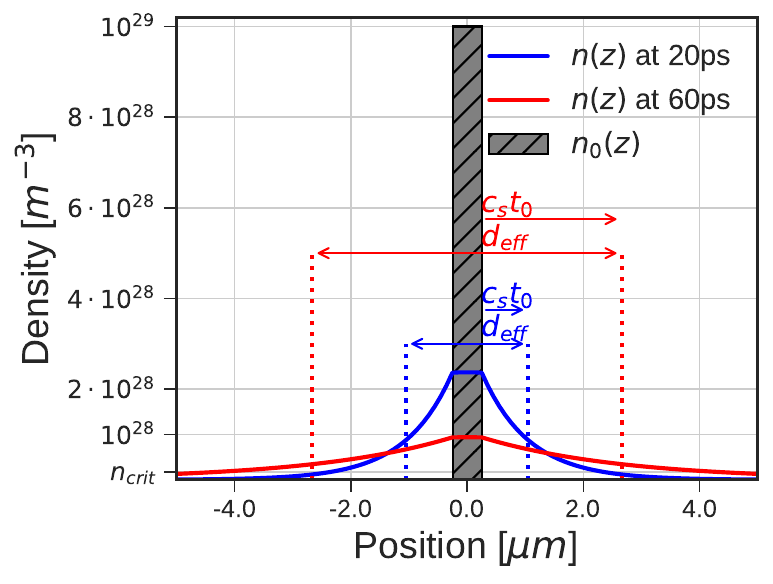}
    \vspace{-2mm}
    \caption{The electron density pre ($n_0(z)$) and post ($n(z)$) expansion for laser parameters of minimum peak intensity $I_0 = 10^{18}$ \unit{\watt \per \centi \meter \squared}, minimum contrast $\kappa = 10^{-7}$, and minimum thickness $d_0 = $ \SI{0.5}{\micro \meter} is plotted. The position $z$ is the distance from the center of the target. In blue and red, $n(z)$ is plotted using $t_0 = 20$~ps and $t_0 = 60$~ps respectively for the time elapsed since the arrival of pre-pulse.}
    \label{fig:density}
\end{figure}

In two experimental studies, \citet{Morrison_etal2018} and \citet{Loughran_etal2023}, the highest proton energy results did \textit{not} occur when the target was placed at peak focus.  In both papers, this result was attributed to pre-pulse effects. Our model includes a prescription to include this effect.

The \citet{Mora_2003} model (which considers the 1D expansion of a hot plasma into a vacuum) for predicting the peak proton energy can also be used to estimate the pre-expansion of the target due to a presence of a pre-pulse. 
For simplicity, we treat the pre-pulse as a spike in intensity that occurs 60~ps before the arrival of the main laser pulse. 
During this 60~ps, the target can expand because of the energy deposited by the pre-pulse. In our investigations, we kept the time delay fixed but we varied the ``contrast", which is the ratio between the intensity of the pre-pulse and the peak intensity of the main pulse (i.e. $\kappa = I_\text{pre} / I_\text{main}$). 

As discussed in \citet{Mora_2003}, 1D plasma expansion produces an exponential decay profile in target density extending from the edges of the target with a characteristic length of $c_s t_0$, where $c_s$ is the speed of sound within the target plasma and $t_0$ the time elapsed since the arrival of the pre-pulse, as depicted in \autoref{fig:density}. We determine an effective thickness of the target as it expands according to the following equation:
\begin{equation}
    \label{eqn:effThickImplemented}
    d_{\text{eff}} = d_0 + 2 c_s t_0
\end{equation}
where $d_0$ is the initial thickness. This thickness is smaller than the thickness one would obtain from measuring the distance from the critical density surface on one side of the target to the other.

We assume that the speed of sound within the plasma is given by
\begin{equation}
    c_s = \sqrt{\frac{Z_i k_B T_e }{m_i}}
\end{equation}
where $Z_i$ represents the ion charge, $T_e$ the electron temperature, and $m_i$ the mass of the ion. Assuming a hydrogen target makes $Z_i = 1$ and $m_i$ equal to the mass of a proton. Furthermore, the thermal energy absorbed by the target electrons is taken to be proportional to the laser intensity, $k_B T_e \propto I$, where $k_B$ is Boltzmann's constant. This implies that the pre-pulse heating happens before the target can expand and quickly enough that the target does not cool by radiative processes. We assume that an intensity of $10^{12}$ \unit{\watt\per\centi\meter\squared} produces electron temperatures of 50~eV. Other intensities produce electron temperatures in proportion to those values.

As previously mentioned, experimental studies \cite{Morrison_etal2018, Loughran_etal2023} noticed that the best proton acceleration occurred when the target was placed many microns away from the laser peak focus. There are two effects in our model that produce this behavior: defocusing and pump depletion. Regarding defocusing, our model includes a prescription for reducing the laser intensity if the target is off of peak focus. Specifically, we assume that the laser is coming to focus as a perfect Gaussian beam and that placing the target off of peak focus will reduce the main pulse and pre-pulse intensity according to the Gaussian beam formula. 
Because the pre-pulse intensity is reduced by the same factor, when the target is off of peak focus, the target expands less, so the main pulse sees a thinner target, which can potentially help increase the maximum proton energy. 

%This may be worth rewriting: if I remember correctly, defocusing never causes increased proton energy; increasing the intensity always increased energy unless doing so would take the target below critical density, which we did not observe in our data set.  It is true that defocusing made the energy falloff from peak intensity less steep, but defocusing alone without pump depletion never improved energy.

Pump depletion can also affect the peak proton energy because the main laser pulse has to travel through an extended pre-plasma to reach the target. 
We describe a simple model based on arguments in \citet{Decker} to capture this effect. \citet{Decker} considers laser beams that are less tightly focused than the typical beams in proton acceleration experiments, so the applicability of their arguments to our case may be limited, but we include them in order to provide a qualitative description of pump depletion motivated by physical considerations.

We assume that the beam travels through vacuum except for the exponential pre-plasma that extends from the target \cite{Mora_2003}. As depicted in \autoref{fig:density}, the distribution of the pre-plasma is symmetric, and we assume this to be true regardless of the position of the target relative to the focus. 
\citet{Decker} describes pump depletion as an ``etching" process where traveling through the plasma causes the edge of the wavefront to recede at a speed given by the ``etching velocity," $v_\text{etch} = (\omega_{p,e}/\omega)^2 c$. The plasma frequency is given by $\omega_{p,e}^2 = n_e e^2 / m_e \epsilon_0$ where $n_e$ is the electron density, so the etching velocity is different depending on the position in the plasma. We can integrate the etching velocity to obtain the ``etching length" which is given by $L_{\rm etch} = \int v_\text{etch} dt$. By applying this equation to our exponential model of the pre plasma, we find that:
\begin{equation}
    \label{eqn:etchLength}
    L_{\text{etch}} = \frac{e^2 n(0) c_s t_0}{\epsilon_0 m_e \omega^2} \left( \exp{\left(-\frac{z_0}{c_s t_0}\right)} - \exp{\left(-\frac{z_f}{c_s t_0}\right)} \right)
\end{equation}
where $z_0$ is the position of critical density (i.e. the barrier between the under- and overdense plasma), $z_f$ is the position of the front of the expanded target (i.e. the barrier between the underdense plasma and the vacuum), and $n(0)$ is the maximum electron density post expansion (see \autoref{fig:density}). These positions are with respect to the edge of the pre-expanded target.

The pulse duration of the laser is shortened by dividing the etching length, $L_\text{etch}$, by the speed of light, $c$
\begin{equation}
    \label{eqn:tau_eff}
    \tau_{\text{eff}} = \tau_\text{laser} - \frac{L_\text{etch}}{c}.
\end{equation}
If $L_\text{etch}/c$ exceeds $\tau_\text{laser}$, then the entire pulse has been dissipated in the underdense plasma. If instead $L_\text{etch}/c$ does not exceed $\tau_\text{laser}$ then we assume that the remaining energy in the laser is the original laser energy times $\tau_\text{eff} / \tau_\text{laser}$

\begin{figure}
    \centering
    \includegraphics[width=3in]{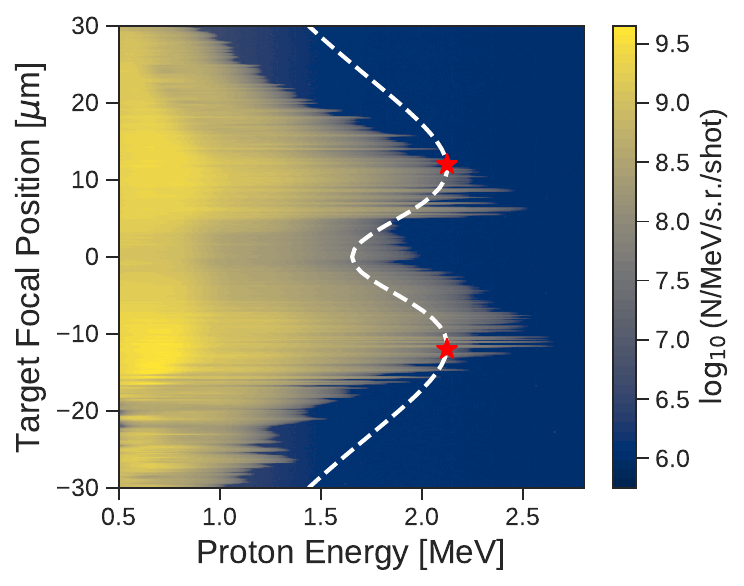}
    \vspace{-4mm}
    \caption{The maximum proton energy as a function of target focal position for laser parameters of maximum peak intensity $I_0 = 10^{19}$ \unit{W . cm^{-2}}, minimum contrast $\kappa = 10^{-7}$, and minimum thickness $d_0 = $ \SI{0.5}{\micro \meter} is plotted (white, dashed). This prediction is overlaid on top of the proton energy measurements from \citet{Morrison_etal2018} (their Figure 3b) collected from a target position scan in the laser propagation direction. The red stars indicate the optimal focal positions ($\pm$\SI{12}{\micro \meter}), where the highest maximum proton energy (\SI{2.12}{\mega \electronvolt}) in the synthetic data spectrum is observed. \href{https://iopscience.iop.org/article/10.1088/1367-2630/aaa8d1}{MeV proton acceleration at kHz repetition rate from ultra-intense laser liquid interaction} © 2018 by John T Morrison et al. is licensed under  \href{https://creativecommons.org/licenses/by/3.0/}{CC BY 3.0}}
    \label{fig:dip}
\end{figure}

Using this model, we produced a focal depth vs proton maximum kinetic energy profile seen in \autoref{fig:dip}, showing maxima in the proton energy off of peak laser focus. \autoref{fig:dip} specifically highlights predictions for a target \SI{0.5}{\micro \meter} thick, an \SI{800}{\nano \meter} wavelength, \SI{40}{\femto \second} laser pulse (before etching), a laser intensity of $10^{19}$ \unit {\watt \per \centi \meter \squared} at peak focus, a laser spot size of \SI{1.5}{\micro \meter}, and an initial target electron density of $10^{29}$ \unit{\meter^{-3}}. In \autoref{fig:dip} this prediction is compared directly with experimental measurements from \citet{Morrison_etal2018}. \autoref{fig:dip} shows that our theoretical model qualitatively matches the experimental results, which is sufficient for the purposes of this paper.

%The features displayed in \autoref{fig:dip} make more interesting the task of training machine learning models on this physical model and using them for optimization problems. The ML models must be able to reproduce the ``dip" at maximum proton energy to ascertain optimal input conditions to produce proton energies approaching \SI{2}{\mega \electronvolt}. This feature was not included in our earlier work which used a synthetic data model that always produced optimal proton energy at peak focus \cite{Desai}.

\subsection{Range of Synthetic Data Generated} \label{sec:range}

Our synthetic data set is designed to mimic an ultra-intense laser system with \SI{40}{\femto \second} pulses, pulse energies between \SI{1.41}{\milli\joule} and \SI{14.14}{\milli \joule}, and a peak focus spot size of \SI{1.5}{\micro \meter}. These values are similar to the laser system described in \citet{Morrison_etal2018}. As discussed in the previous section, the on-target laser intensity in our synthetic data depends on the position of the target. If the target is placed at peak focus the laser intensity is near $10^{19}$~\unit{W . cm^{-2}} for the highest pulse energy we considered.

Experimentally, there are many different ways to explore the parameter space. We generated a grid of input points using the ``meshgrid" function of the NumPy \cite{numpy} package.  This grid consists of target conditions that are experimentally feasible (e.g. \cite{Morrison_etal2018}), varying target thickness from 0.5 to \SI{5.0}{\micro \meter} in steps of \SI{0.5}{\micro \meter}, target focal position from -30 to 30~\unit{\micro \meter} in steps of \SI{1.0}{\micro \meter}, peak laser intensity from $10^{18}$ to $10^{19}$~\unit{\watt \per \centi \meter \squared} in steps of $1.8 \times 10^{17}$~\unit{\watt \per \centi \meter \squared}, and the contrast from $10^{-7}$ to $10^{-6}$ in steps of $1.8 \times 10^{-8}$, creating a training data set with 1,525,000 points.  Noise was added to the model in a manner that will be described in the next subsection. For parameter choices where the proton energies were very low, the maximum and average proton energies were set to 1~$\unit{keV}$ and the total proton energy set to 1~$\unit{nJ}$ before applying any noise. These very low proton energy conditions can happen, for example, when the pre-pulse expands the target so much that the effective target thickness is very large.
We apply these lower bounds because we are uninterested in proton energies below this level. Likewise, it is unnecessary for the ML models to be accurate below these levels.

For the testing data set -- unlike the training data set --  we randomly sampled the entire parameter space, generating 250,000 points within the same minimum and maximum bounds as the training set. This approach ensures that the ML models are often being evaluated for parameters that were not included in the training set. For example, the training set only contains half-integer thicknesses from 0.5 to \SI{5.0}{\micro \meter}, but the testing set can include targets of any thickness between these bounds. This approach mimics the experiment because the exploration of parameter space is always limited and optimum conditions for proton acceleration could easily occur outside of the experimental conditions that have already been explored. In Appendix~\ref{ap:campaigns}, we briefly describe results from an even more realistic scheme for exploring the parameter space.

The testing data set did \textit{not} include added noise, whereas the training set does include added noise. In this way, the testing set performance represents the ability of the models to average out the shot-to-shot variations of the laser system and extrapolate beyond the provided input parameters.

\subsection{Noise model}
\label{sec:noise}

Ultra-intense laser experiments typically involve significant shot-to-shot variations, so we added noise to all three output quantities of the training set -- maximum proton energy, total proton energy, and average proton energy. As in our earlier work \cite{Desai}, we added noise by sampling from a log-normal distribution using each prediction as the mean. Importantly, this approach assumes that experiments with larger laser energy and intensity will involve greater shot-to-shot fluctuations. We follow the same conventions as \citet{Desai} when we refer to the percentage of ``Gaussian" noise added.  To evaluate how different noise levels affect the performance of the machine learning models, training sets with noise levels ranging from 0\% (noiseless) to 30\% Gaussian noise were generated.
The interested reader can see our exact implementation of this noise model by viewing the source code \cite{desai_2024_zenodo} included with this publication.

\section{Machine Learning Methods}
\label{sec:ml}

We use the synthetic data to train three different machine learning (ML) models -- a Neural Network (NN), Stochastic Variational Gaussian Process Regression (SVGP), and Polynomial Regression (POLY). The polynomial regression is by far the simplest model that we tested and contains the fewest free parameters. Some would argue that the polynomial regression is sufficiently simple that it should not be categorized as a machine learning model, or that SVGP is a statistical rather than a machine learning model. We acknowledge this point of view and we clarify that we are using a broad definition of the term machine learning when we label all three of these methods as machine learning models.

The ML models were trained on the available hardware through the Ohio Supercomputing Center (OSC) resources. The ML models' performance was evaluating using both a CPU and GPU at different times during testing; an AMD EPYC 7643 was used as the CPU, whereas an Nvidia A100 80 GB was used as the GPU. 

As mentioned, there are four independent variables in the data set -- target thickness, intensity, focal distance, and contrast. We apply a natural logarithm to the intensity and the contrast, as well as the three outputs of the model: maximum, total, and average proton energy. Then, we apply min-max scaling which is a linear re-scaling that casts the minimum and maximum values to 0 and 1 respectively for each of the inputs and outputs in the dataset. This scaling was determined only from the inputs and outputs from the training set. As in our earlier work \cite{Desai}, we correct for bias introduced by the log-scaling of the outputs following arguments in \citet{miller_1984}.

As the information relevant to the NN and Gaussian Process remains the same as our prior paper, see \citet{Desai} for an in-depth explanation of the properties of each model. In the following subsections, we only highlight explicit differences from the prior implementations, especially with regards to scaling the models to allow for training sets in excess of 1 million data.

\subsection{Polynomial Regression}
\label{sec:Poly}

The POLY is the simplest and least computationally expensive of all models evaluated. For this reason, GPU computations were entirely unnecessary.  Training a POLY on even the largest data sets took less than a minute, even when only trained on a single CPU core.

The Scikit-Learn library was used to implement the POLY model.  To fit a degree-$p$ regression, the implementation creates polynomial features containing all products of input features up to degree-$p$ to use as independent variables (e.g. a degree-2 regression on data with 2 input features $x_1$ and $x_2$ would fit the output data to $y = \beta_0 + \beta_1 x_1 + \beta_2 x_2 + \beta_3 x_1^2 + \beta_4 x_2^2 + \beta_5 x_1 x_2$).  Each output feature was fitted to an independent set of weights to allow for varied predictions for each output dimension.  The fitting process then fits a ridge regression from each output dimension to its polynomial function of input features, which includes a penalty for weight terms that allows for more general trends in the data to be learned.

We found that the optimal performance for the POLY was achieved with a degree-7 polynomial with a regularization factor of $10^{-3}$ (Appendix~\ref{ap:hyper}).

\subsection{Neural Network}
\label{sec:nn}

\begin{figure}
    \centering
    \includegraphics[width=3.5in]{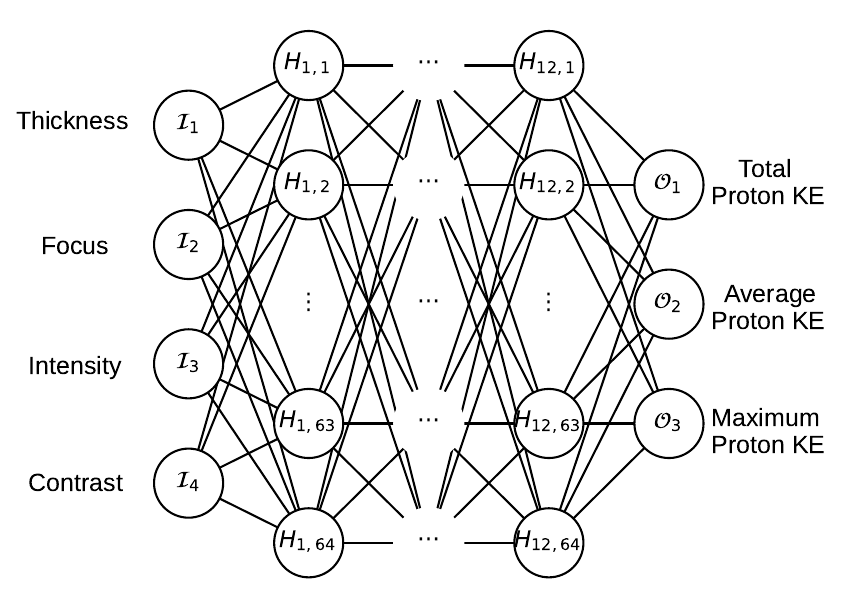}
    \caption{Architecture of the neural network model. The 4 inputs (target thickness, target focal position, peak intensity, and contrast) and three outputs (maximum, total, and average proton energy) are fully connected by 12 hidden layers of 64 neurons each. Vertical and horizontal dots represent the neurons not shown and each neuron has connections to every other neuron located in the adjacent layers.}
    \label{fig:nn_arch}
\end{figure}

To create a NN, we used the Skorch \cite{skorch} wrapper of the PyTorch library, enabling our model to benefit from the high-level Scikit-Learn API and the performance of PyTorch.  For this regression task, we used a fully-connected network (\autoref{fig:nn_arch}).  As part of the hyperparameter optimization process, we allowed the model to assume variable architecture: different numbers of nodes and layers, activation functions, application of the batch normalization function, learning rates, and back-propagation schemes were tested as discussed in Appendix~\ref{ap:hyper}.

We found an optimal model architecture to be a network with 12 hidden layers, 64 nodes per layer, with batch normalization and the nonlinear ``Leaky ReLU" activation function applied between layers (Appendix~\ref{ap:hyper}).  We trained the neural network using the ``Adam" scheme for back-propagation as described in \citet{kingma2017adam}, which outperformed alternative methods (such as stochastic gradient descent) during optimization.

Much consideration was devoted to minimizing the possibility of over-fitting the NN during training.  A network with $\ell$ layers and $n$ nodes per layer has $\mathcal{O}(n^2 \ell)$ trainable parameters (due to the connections between all nodes in adjacent layers), so large models can quickly gain more trainable parameters than there are data, restricting the degrees of freedom available to train the model.  To combat this issue, the model was trained with early stopping implemented: once a certain number of training epochs passes with no reduction in loss on the validation set, the model exited the training process early. The specific number of epochs without improvement that triggers the early stopping is called the ``patience".  

In addition, we considered methods for avoiding the model parameters being stuck in a local minimum for loss as opposed to a global minimum.  To enhance the Adam method's natural ability to skip over local minima, we used an exponential learning rate scheduler while training the model.  This choice allowed us to set the initial learning rate to a high value, which was then reduced by a multiplicative factor $\gamma$ after each epoch during the training process.  This approach allows the training process to skip over local minima while reducing the coarseness of the search to properly home in on optimal parameter values for achieving minimum loss. As discussed in Appendix~\ref{ap:hyper}, we found that an initial learning rate of 0.01, a patience of 10 epochs, and a decay factor of $\gamma$=0.90 produced the best results.

\subsection{Gaussian Process Regression}
\label{sec:gp}

We direct the reader to \citet{Schulz2017ATO} for an excellent introduction to GPR. To accelerate the computations with GPU, we used the Skorch wrapper of the GPyTorch library \cite{gardner2018gpytorch} version 1.11.  

A well known limitation of the Gaussian process is the unfavorable scaling, with $n$ data points requiring $\mathcal{O}(n^2)$ memory storage and $\mathcal{O}(n^3)$ computations \citep{wang2019exact}.  These limitations became apparent when expanding the size of the training set, as the GPU would run out of available VRAM once the training set exceeded about 40,000 data points in size. There are a few methods for reducing the innate complexity of the GPR. The first of which, proposed in \citet{wang2019exact}, is to use kernel partitioning techniques to drastically reduce the amount of necessary computations, allowing for Gaussian Processes to be trained on data sets exceeding one million points in size.  The other option is to use a Stochastic Variational Gaussian Process (SVGP) as in \citet{svgp} (2013), which assumes a variational distribution over some amount of inducing points, restricting the training data to a representative subset.  This technique also allows Gaussian Processes to be trained on data sets on the order of one million data points, and the data can be split into smaller ``mini-batches" (as in the training process of the NN) to aid computational efficiency.

%For our implementation, we initially used the KeOps library \citep{KeOps} to speed up calculations.  This allowed the GPR to handle larger training sets, but once the training set exceeded $\sim 100,000$ data points, an unavoidable CUDA-level memory error would be encountered, cutting short the training process.  As such, the kernel partitioning approach was abandoned in favor of the SVGP, which was able to utilize the entire training set.

For our implementation, we used SVGP. 
As is standard for SVGPs, we used the variational evidence lower bound (ELBO) introduced in \citet{scaling} as the loss function to minimize during training.  As discussed in Appendix~\ref{ap:hyper}, we found optimal results with 8 latent functions and 2000 inducing points, trained using the Adam method in batch sizes of 1024 with a learning rate of 0.01. It is important to note that increasing the number of inducing points beyond 2000 significantly increased training time, putting a soft practical limit at 2000 inducing points.  It is possible that better results would be achievable with more inducing points, but this would come at the cost of significantly increased training cost in terms of time and resources.

\subsection{Hyperparameter Optimization} \label{sec:hp_opt}

As has already been mentioned, each model contained several hyperparameters that are selected by the user rather than updated as part of the training process. All of the models require hyperparameters, including the polynomial regression. As described in Appendix~\ref{ap:hyper}, to ensure that each model contained an optimal selection of hyperparameters, we used a grid search to evaluate model performance across several combinations of hyperparameters when possible, restricting our grid as we homed in on the optimal model architecture \citep{Geron_2023}.  This method splits the training set into $k$ cross-validation splits.  Then it trains models with each possible combination of hyperparameters along a specified grid $k$ times, using each split as a validation set once per combination and the remaining data as a training set.  This process ensures that each model can have the optimal structure for our specific case without over-representing any one region of the data.

For an in-depth discussion of the hyperparameter selection process and results, see Appendix~\ref{ap:hyper}.

\section{Training Time and Accuracy}
\label{sec:results}

\begin{figure}
    \centering
    \includegraphics[width=2.75in]{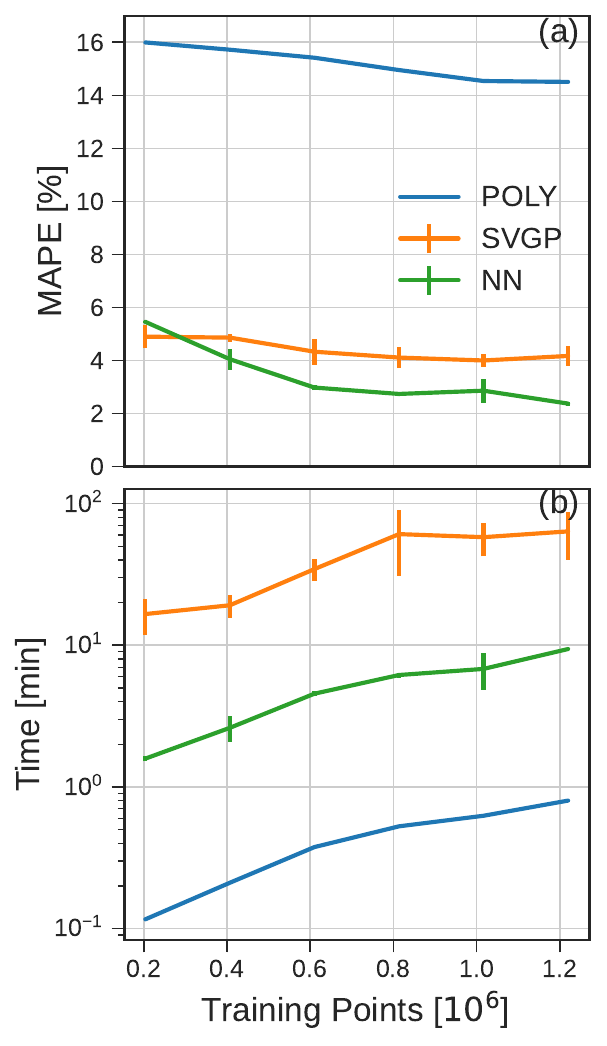}
    \vspace{-0.5cm}
    \caption{Model training results using data with 10 \% added noise. Testing MAPE (a) is plotted for the three ML models against the number of training points and averaged between results for maximum, average, and total proton energy. The training time (b) of the ML models in minutes is plotted on a logarithmic scale. The vertical bars are standard deviations computed from running the training splits 3 times with different seeds to control the data splitting and random parameter initialization of the NN and SVGP models.}
    \label{fig:train_time_accuracy}
\end{figure}

\begin{figure}
    \centering
    \includegraphics[width=2.75in]{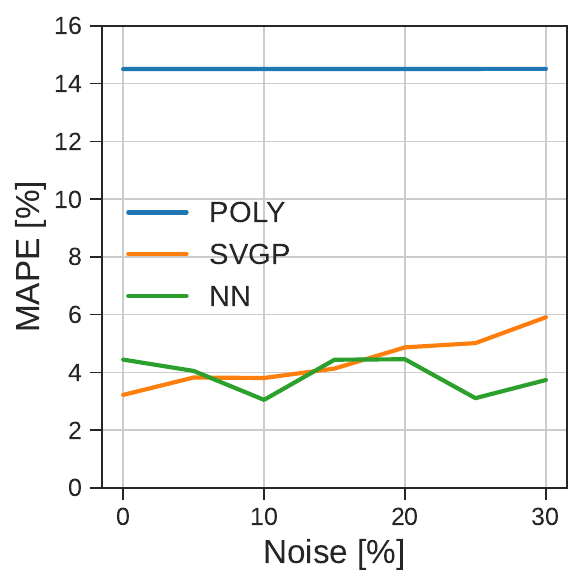}
    %\vspace{-0.5cm}
    \caption{Testing MAPE is plotted against different levels of gaussian noise using the full training dataset for the three models with the three output energy results averaged.}
    \label{fig:noise_accuracy}
\end{figure}

\autoref{fig:train_time_accuracy}a presents results for the mean absolute percentage error (MAPE) averaged between the three outputs for the three ML models discussed in Section~\ref{sec:ml}: the polynomial ridge regression of degree 7 (POLY), the 12x64 architecture neural network (NN), and the stochastic variational gaussian process (SVGP). Each model is trained on data with 10\% Gaussian noise. The MAPE is computed with respect to the testing dataset described in \autoref{sec:range} without added noise. The number of training points was varied by randomly sampling 80\% of the full 1,525,000 point dataset for training with reserving the other 20\% for validation used during the NN and SVGP training. The NN has a lower percentage error than the SVGP for most of the training splits. The NN and SVGP models are much more accurate than the POLY model. 

The training time is plotted in \autoref{fig:train_time_accuracy}b on a logarithmic scale for the three models. Generally we find that the POLY was fastest, with the NN requiring an order of magnitude more time and SVGP requiring another order of magnitude more time than the POLY. 

Figures~\ref{fig:train_time_accuracy}a and \ref{fig:train_time_accuracy}b indicate that the accuracy of the POLY model was an order of magnitude worse than the other two ML models, but the execution time was at least an order of magnitude faster for POLY than the other ML models, despite only using one core of a CPU while the other two ML models are using a GPU. In both \autoref{fig:train_time_accuracy}a and \autoref{fig:train_time_accuracy}b, values were averaged over three runs to mitigate any bias due to a particular parameter initialization or data split.

\autoref{fig:noise_accuracy} shows the testing MAPE plotted against the level of gaussian noise added to the training set, fixing the number of data points to the full set. The accuracy of both the NN and POLY models appear to be robust to increasing the noise level. The accuracy of the SVGP model shows some weak dependence on the noise level.

\section{Using Trained Models to Evaluate an Objective Function}
\label{sec:opt}

\begin{figure*}
    \centering
    \includegraphics[width=6in]{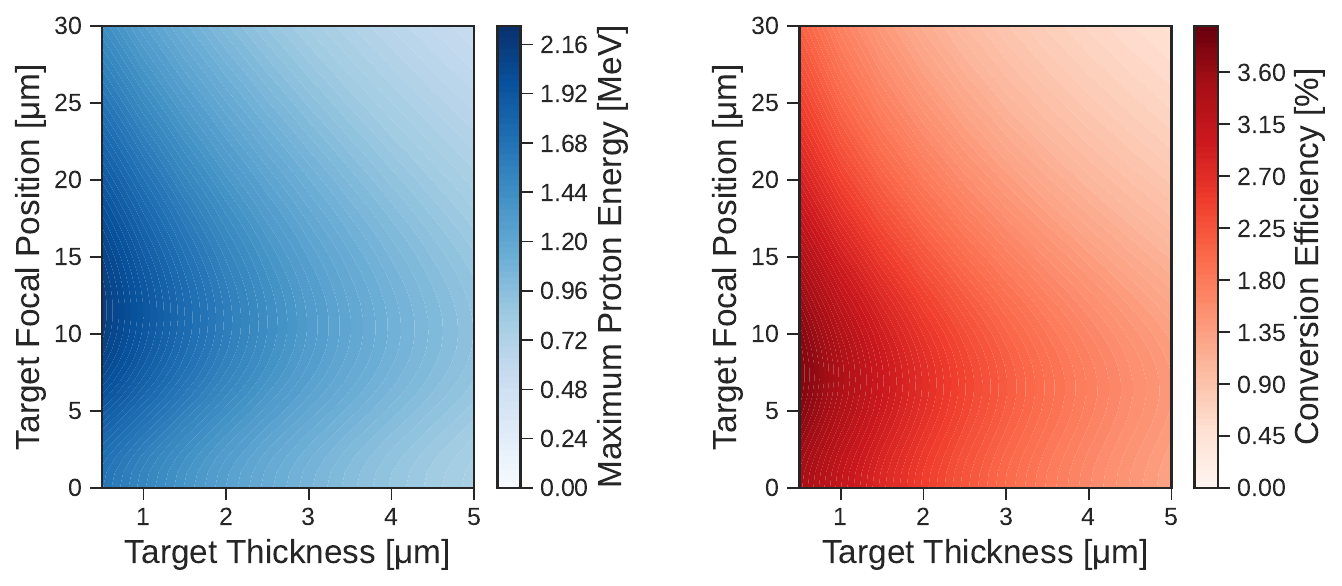}
    \vspace{-0.4cm}
    \caption{Colormaps in the 2D parameter space of target thickness and target focal position that display (left panel) the maximum proton energy (i.e. energy cutoff $KE_{\rm c}$) and (right panel) laser to proton energy conversion efficiency $\eta_{p}$ as calculated from the modified Fuchs et al model. These plots were generated assuming 14.14~mJ of laser energy and a pre-pulse contrast of $10^{-7}$.}
    \label{fig:energy_efficiency}
\end{figure*}

A surrogate model that accurately predicts the proton energy spectrum from the laser input conditions would be very useful for obtaining the independent variables (i.e. laser and target conditions) that enhance or optimize the proton energy distribution in some way. With this in mind, we used trained models to evaluate an objective function. This is a similar objective function to Equation 3 of \citet{Desai}. For each of the three trained models we brute force search for the minimum of this function
\begin{eqnarray}
     f(KE_{\rm c},\eta_{p}) = \frac{|KE_{\rm c} - KE_{\rm c,goal}|}{1 \text{MeV}} \beta - 100 \eta_{p} (1-\beta) \label{eq:function}
\end{eqnarray}
where $KE_{\rm c,goal}$ is the user-requested ``cutoff" proton kinetic energy which can also be described as the maximum kinetic energy of the TNSA proton energy distribution, $KE_{\rm c}$ is the predicted cutoff proton kinetic energy from the model for a specific set of independent variables, $\eta_p$ is the conversion efficiency from laser energy to proton energy, and $\beta$ is a unitless parameter between 0 and 1 that sets the relative importance of matching to the user-requested cutoff energy versus increasing the conversion efficiency ($\eta_p$). Because the trained models can quickly make predictions, we can evaluate \autoref{eq:function} for many different sets of laser and target parameters and find the values that minimize this function for a particular choice of $KE_{\rm c,goal}$ and $\beta$. In a moment we will show results from setting $KE_{\rm c,goal} = 1$~MeV and exploring a few different values of $\beta$ for each ML model. The training data includes 30\% Gaussian noise. Because the ML models were trained on data from an analytic model based on \citet{Fuchs2005}, we can look for optimum conditions using the same analytic model but without any noise added in order to find the true laser and target parameters that minimize \autoref{eq:function} for $KE_{\rm c,goal} = 1$~MeV and different values of $\beta$. For reference, the predictions from this analytic model without any noise are shown in \autoref{fig:energy_efficiency}. This is the same analytic model that was used to produce the model predictions in \autoref{fig:dip} earlier.

\begin{figure*}
    \centering
    \includegraphics[width=6.5in]{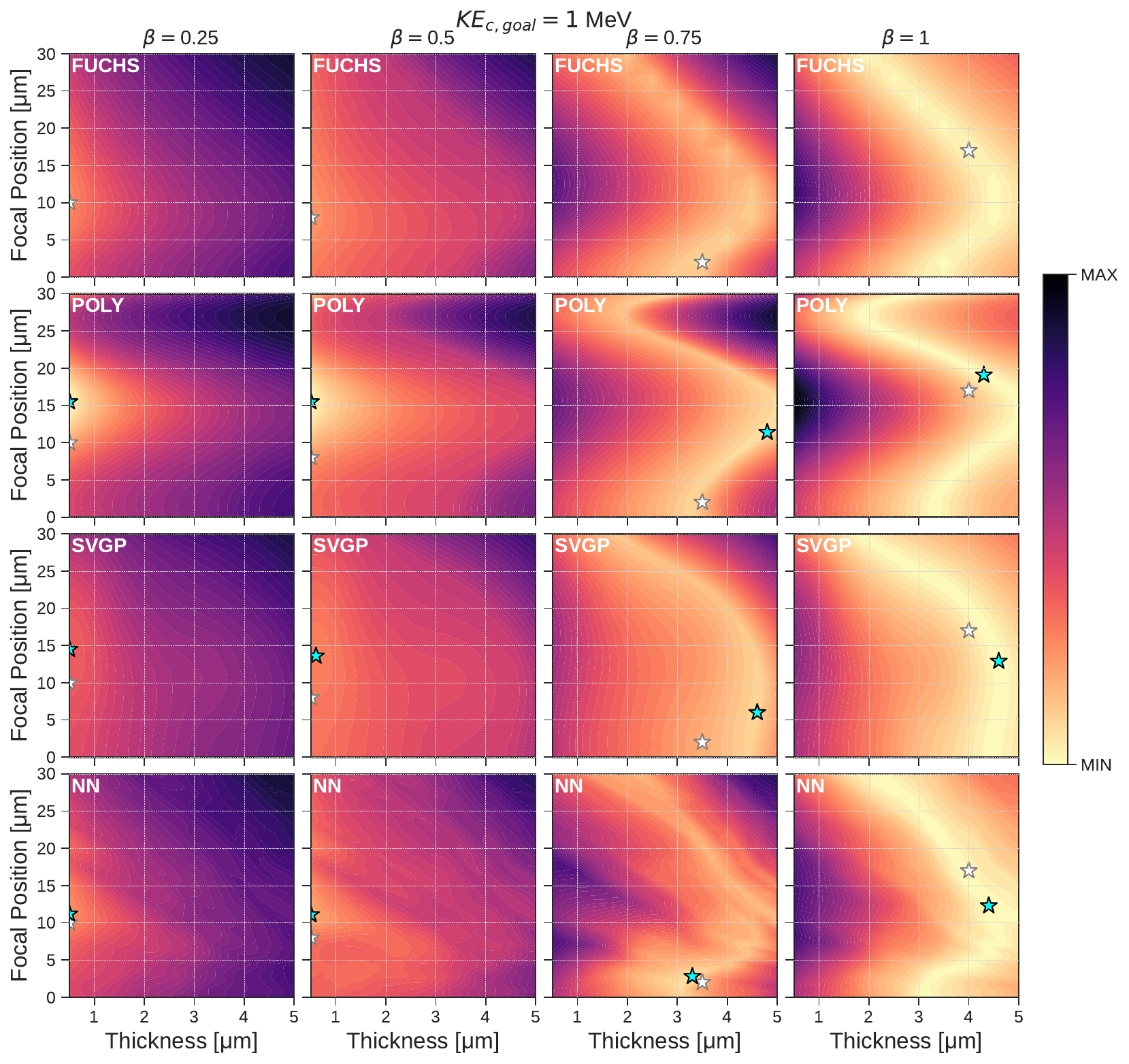}
    \vspace{-0.4cm}
    \caption{Colormaps that show estimates of \autoref{eq:function} assuming $KE_{\rm c,goal} = 1$~MeV for the three ML models (NN, POLY and SVGP) and the modified Fuchs et al. model dataset with no added noise (FUCHS). The modified Fuchs et al. model with added 30\% Gaussian noise was used to produce the training data for the ML models.
    For each $\beta$ value (i.e. each column), the same color levels are used in order to facilitate comparison between the models. A cyan colored star is placed at the location where each ML model predicts a minimum value for \autoref{eq:function} which can be compared to the analytic model prediction indicated by a white star.}
    \label{fig:obn_fn_model}
\end{figure*}

In performing this comparison, which is shown in \autoref{fig:obn_fn_model}, for simplicity we fix the laser energy to 14.14~mJ and the contrast to ($10^{-7}$) in our training set. Then, we vary the target thickness and target focal position. \autoref{fig:obn_fn_model} was created by evaluating \autoref{eq:function} on a 2D grid of points with thickness ranging from 0.5 to 5 $\mu$m (in 0.1 $\mu$m steps) and target focal position ranging from 0 to 30 $\mu$m (in 0.1 $\mu$m steps) for a total of 13846 points. The time taken to perform the inference step of taking a trained model and evaluating it on these 13846 points is only a few seconds or less using just CPU inference with the NN and POLY. The SVGP takes considerably longer (minutes). Note that the step sizes are five to ten times larger than what was used in the dataset generation described in Section~\ref{sec:range} so that we are interpolating between points not originally seen by the trained models. \autoref{fig:obn_fn_model} is similar to Figure~6 from \citet{Desai}, but the optimum conditions for proton acceleration often occur away from the peak focus (i.e. at non-zero values for the focal position), as was highlighted earlier in \autoref{fig:dip}. 

The optimum conditions (i.e. the conditions that minimize \autoref{eq:function}) are indicated with a star. The upper panels of \autoref{fig:obn_fn_model} show the results from the analytic model whose optimum conditions are indicated by a white star. These panels should be regarded as the true distribution of \autoref{eq:function} versus thickness and focal position. The other panels show the estimated distribution of \autoref{eq:function} versus thickness and focal position from the different ML models whose optimum conditions are indicated by a cyan star. The white star from the analytic model is displayed for comparison.

\setlength{\tabcolsep}{4pt} % Default value: 6pt
\renewcommand{\arraystretch}{1.5} % Default value: 1

\begin{table}
\centering
\begin{tabular}{|c||c|ccccc|}
\hline
   &$\beta$   & 0        & 0.25     & 0.5      & 0.75     & 1        \\ 
 \hline
 & POLY & 0.329   & 0.223  & 0.121   & 0.057 & 0.124  \\
 RMSE & SVGP & 0.143  & 0.096 & 0.062 & 0.065 & 0.101  \\
 &NN   & 0.065 & 0.052 & 0.042 & 0.038 & 0.042 \\
 \hline 
 & POLY & 4.5 & 5.5  & 7.5     & 9.489  & 2.121 \\
 $\Delta_\text{opt}$ [$\mu$m] &SVGP & 3.5 & 4.5  & 5.601 & 4.148  & 4.144 \\
  &NN   & 0.4 & 1.2  & 3.1     & 0.825 & 4.717 \\
 \hline
\end{tabular}
\caption{Comparison metrics evaluated from \autoref{fig:obn_fn_model}. The RMSE row shows the root mean squared error between the colormap values of the ML models and the analytic model for each value of $\beta$. The $\Delta_\text{opt}$ row calculates the Euclidean distance between the predicted optimum and true optimum (i.e. distance in $\mu$m between the cyan and white stars in \autoref{fig:obn_fn_model}).}
\label{tab:opt_results}
\end{table}

In \autoref{fig:obn_fn_model}, one can see that for $\beta = 0.25$, the best (i.e. globally minimum) region is at 0.5~$\mu$m thickness and 10~$\mu$m focal position, which correspond to regions with higher $\eta_{p}$. The POLY and SVGP models predict the optimal conditions to be at a focal position of $\sim15~\mu$m, in contrast to the NN prediction of $\sim11~\mu$m. In this case, the NN more closely matches the true optimum. For high values of $\beta$, the best region is a curve composed of points that closely match $KE_{\rm c}$ to $KE_{\rm c,goal}$. Using the highest value of $\beta = 1$, we see that while the NN looks much less smooth, it fits the overall shape of the underlying Fuchs model better than the other two methods.

To quantify the features of \autoref{fig:obn_fn_model}, we can look to \autoref{tab:opt_results}. We assess the accuracy in the optimal conditions by taking the Euclidean distance in the focal position - target thickness space between the true optimum conditions and the ML predictions. This distance, termed $\Delta_\text{opt}$, shows that (with the exception of $\beta=1$), the NN predicted optimum is closer to the true model than the SVGP or POLY. To assess accuracy in the colormap as a whole, we can take the root mean squared error (RMSE) between the analytic model and the ML model's colormap values which clearly show lower error for the NN in comparison to the SVGP and POLY.

\section{Summary and Conclusions}
\label{sec:concl}

Building off of \citet{Desai}, we performed tests where a physically-based analytic model modified from \citet{Fuchs2005} was used to generate synthetic data and these synthetic data were used to train three different ML models. Two significant improvements over \citet{Desai} are that the analytic model can account for the expansion of the target due to pre-heating from a pre-pulse, and the total number of data points is significantly increased in order to anticipate the data sets that will soon be available from experimental facilities. Adding pre-pulse physics to the model allows for optimal proton acceleration when the target is away from peak focus, as was seen in \citet{Morrison_etal2018} and \citet{Loughran_etal2023}. 

The extra physics coupled to the increased size of the data set means that there are interesting features that the ML models need to mimic. We performed comparisons including a test where the trained ML models were used to infer the ideal laser and target parameters to produce significant numbers of protons up to an energy of 1~MeV. 

We find that the polynomial regression model, which is much simpler than the neural network or Gaussian Process Regression-based models, can follow overall trends but does misrepresent features in the training set. Generally, we find that the polynomial regression model is about an order of magnitude less accurate than the neural network or Gaussian process models, but the training time for the polynomial regression model is much faster than the other ML models by \emph{at least} one order of magnitude which could be an important advantage for some applications.

The Neural Network model was able to learn the complex features of our dataset, and its computational performance was overall quite favorable, requiring several minutes to train on 1~Million training points on one GPU. 

The Gaussian Process Regression-based model was more accurate than the polynomial regression. However, this model was less accurate than the neural network and the training time was significant, requiring of order a half hour to train on 1~Million training points on one GPU.

By performing these tests, we demonstrate an improved framework for testing the suitability of different ML models for use in high repetition rate ultra-intense laser experiments. This framework can be further improved, for example, by adding more physics to the analytic model that was used to generate the synthetic data, by testing more ML models on the synthetic data, and by performing optimization tasks over more than two parameters. Additionally, a hybrid approach involving simulation data (similar to BLAST \cite{Sandberg_2024_ACM}) could augment the model to allow some of our free parameters to be learned. We provide Jupyter notebooks with our Python code in addition to the datasets used throughout this paper \cite{desai_2024_zenodo} to facilitate this.

\section*{Acknowledgments}
Supercomputer allocations for this project included time from the Ohio Supercomputer Center. We acknowledge support provided by the National Science Foundation (NSF) under Grant No. 2109222. Any opinions, findings, and conclusions or recommendations expressed in this material are those of the author(s) and do not necessarily reflect the views of the National Science Foundation, Department of Defense, or Department of the Air Force. This paper has been cleared for public release, clearance \#MSC/PA-2024-0302; 88ABW-2024-1010. CO was supported in summer 2022 by the Air Force Office of Science Research summer faculty program. This work was supported by Air Force Office of Scientific Research (AFOSR) Award number 23AFCOR004 (PM: Dr. Andrew B. Stickrath) and by Department of Energy Award number 89243021SSC000084 (PM: Dr. Kramer Akli). We also thank Enam Chowdhury for insightful conversations.

\section*{Data availability statement}
The data that support the findings of this study are openly
available in zenodo \cite{desai_2024_zenodo} at \url{https://zenodo.org/records/14009491}, reference number \nolinkurl{10.5281/zenodo.14009490}
\appendix

\clearpage
\section{Optimizing Model Parameters}
\label{ap:hyper}

ML models contain many free hyperparameters that need to be adjusted for the specific dataset at hand. As mentioned in Section~\ref{sec:hp_opt}, hyperparameter optimization was done via a grid search to determine the best performing models (without the use of the testing set). We used the \texttt{GridSearchCV} method of Scikit-Learn \cite{scikit-learn} for our purposes.

\begin{figure}[h]
    \centering
    \includegraphics[width=3.25in]{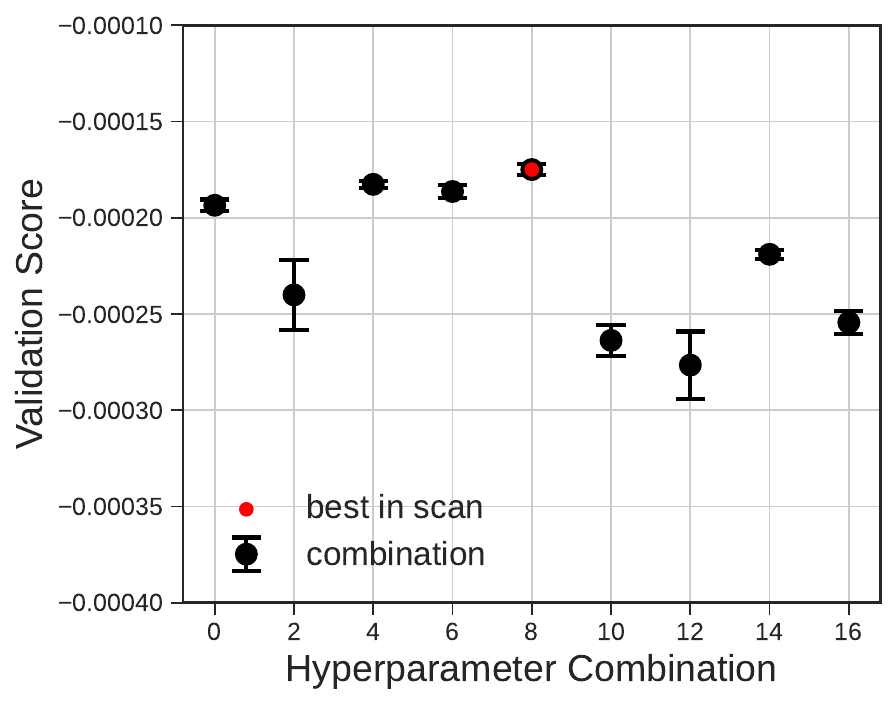}
    \caption{Validation score (using negative mean squared error) with error bars from 3-fold cross-validation plotted for a sampling of the hyperparameters used in a grid search for the NN model. All of the searches were conducted using 1,525,000 point synthetic dataset with 10\% added noise. The following hyperparameters are fixed due to success with prior grid searches (not shown) -- learning rate (0.01), activation function (LeakyReLU), optimizer (Adam), hidden layers (12), neurons per layer (64). The red marker corresponds to the NN model in \autoref{tab:hyperparameters}.}
    \label{fig:nn_hpscan}
\end{figure}

\autoref{fig:nn_hpscan} highlights a sample of the grid searches for the NN model and their validation scores, ranked by the negative mean squared error. The grid searches are done on 3 cross-validation splits, so error bars can also be calculated from the standard deviations between the splits. This particular scan changed the learning rate decay $\gamma$, batch size, and patience. The red-colored point corresponds to the model with the highest score and is what we ultimately chose. Specifics of this model are found in \autoref{tab:hyperparameters}. 

\begin{figure}[h]
    \centering
    \includegraphics[width=3.25in]{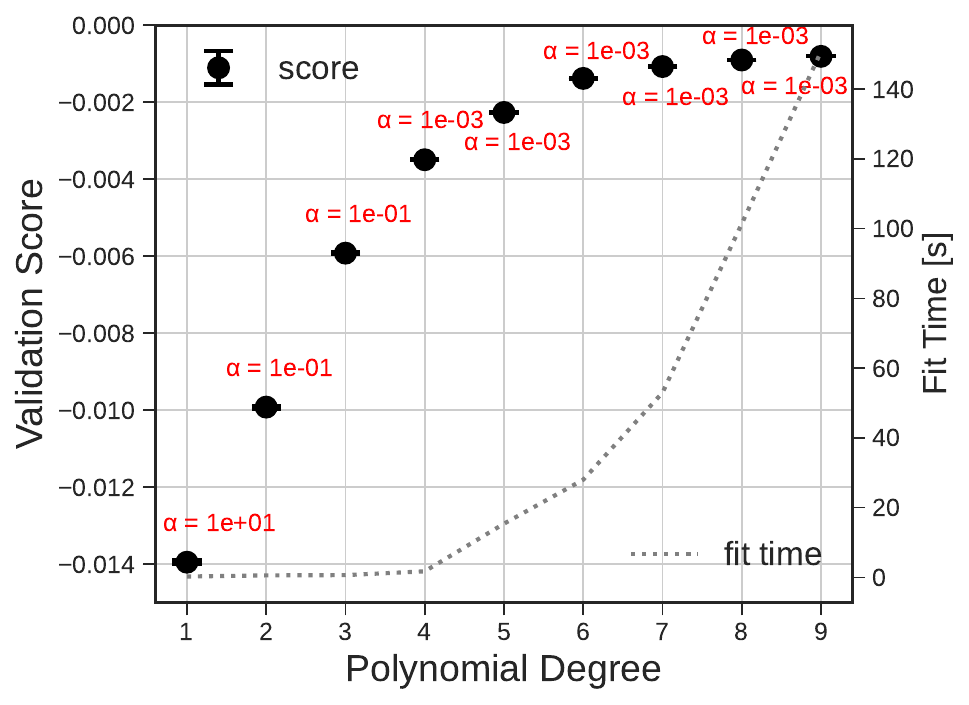}
    \caption{Validation score (using negative mean squared error) from the best ridge regression model at each polynomial degree, tested for regularization parameter $\alpha$ that varied between $10^{-3}$ and $10^{3}$ on a logarithmic scale. Additionally, the fitting time for each degree is overlaid on twin axes.}
    \label{fig:poly_hpscan}
\end{figure}

\autoref{fig:poly_hpscan} shows for the POLY model that the validation score increases to a plateau as the polynomial degree increases. We choose a polynomial degree of 7 for the other investigations in this paper. In \autoref{fig:poly_hpscan}, the regularization parameter $\alpha$ is displayed in red next to the points which varied from $10^{-3}$ to $10^{3}$. We can see that for models of degree greater than 3, the regularization parameter maintains its lowest value of $10^{-3}$ which will make the regularization effect minimal. 

The results for the optimal hyperparameters for each model are summarized in \autoref{tab:hyperparameters}. The SVGP utilized a similar grid search as the NN in a reduced capacity due to the significantly longer run time of the SVGP model.

\begin{table}
    \centering
    \small
    \begin{tabular}{c} 
    {}
    \end {tabular}
    \begin{tabular}{|c|c|c|c|}
    \hline
        Model & Parameters & RMSE & Time (min)   \\
        \hline
        POLY & deg=7, $\alpha=10^{-3}$  & 0.033 & 0.883 \\ 
        \hline
        SVGP & IP=2000, LF=8, LR=$10^{-2}$  & 0.018 & 34.66 \\ 
        \hline
        NN  & BS=$2^{13}$, $\gamma=0.90$, P=10 & &\\
        & LR=$10^{-2}$, 12x64  & 0.013 & 4.93 \\ 
        \hline
    \end{tabular}
    \caption{Optimal hyper-parameters including the root mean square error (RMSE) and mean fit time, determined from a grid search of hyperparameters for the NN, POLY, and SVGP. For the NN, the batch size (BS), learning rate decay ($\gamma$), patience (P), learning rate (LR), and architectures (layers x neurons per layer) were changed throughout the scans. For the POLY, the regularization parameter and degree were varied as seen in \autoref{fig:poly_hpscan}. For the SVGP, the number of inducing points (IP), latent functions (LF), and learning rate (LR) were changed throughout the scans.  See the full results at Zenodo \cite{desai_2024_zenodo}}
    \label{tab:hyperparameters}
\end{table}

\section{Constrained Data Campaigns}
\label{ap:campaigns}

\begin{figure*}
    \centering
    \includegraphics[width=6.5in]{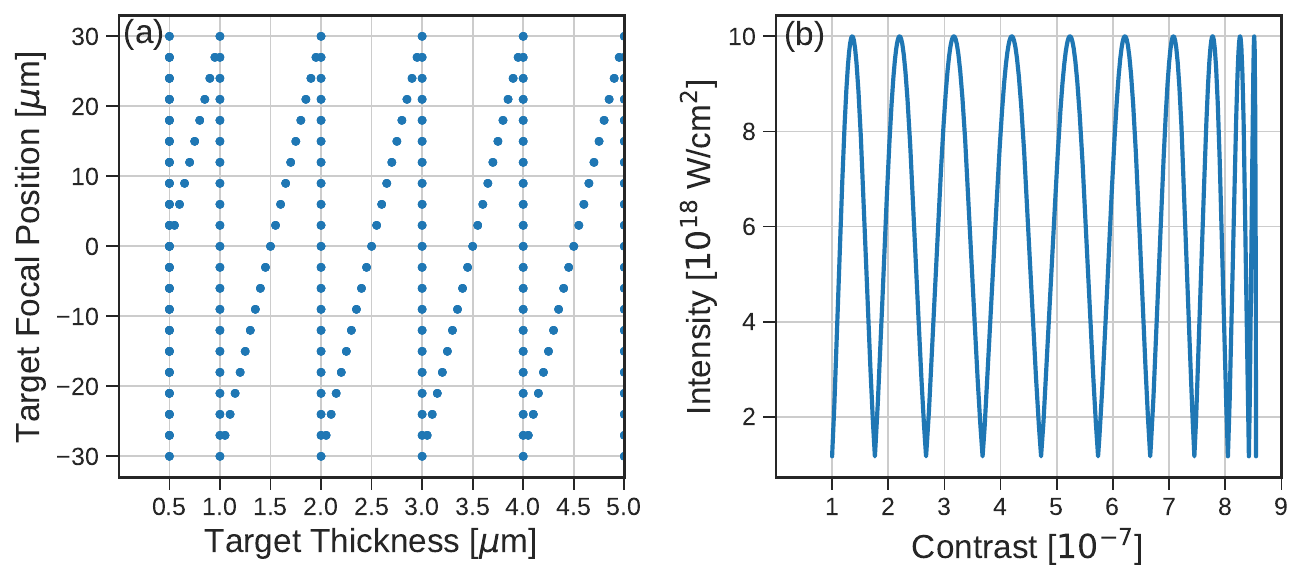}
    \vspace{-0.4cm}
    \caption{Synthetic data was generated in one of two ``campaigns''. In (a) campaign 1, the target focus and thickness is varied in discrete steps and each blue dot varies the laser energy from minimum to maximum. In (b) campaign 2, the depicted intensity and contrast looping is performed for discrete steps in target thickness from $0.5~\mu$m to $5\mu$m}
    \label{fig:zigzag}
\end{figure*}

Earlier in Section~\ref{sec:range}, we describe how synthetic data was generated using a uniform grid in multiple dimensions of laser and target parameters. This scheme was how the primary training set data for the ML models discussed in the main body of the paper was obtained. From an experimental point of view, this approach is not very realistic because a real laser system will scan through the laser and target parameters in a very specific way with specific choices, for example, about which parameters to vary first, while keeping other parameters constant, and which parameters to vary later while keeping other parameters constant. How do these choices affect the accuracy of ML models trained on this data? This is a question that we cannot yet answer conclusively, but we include in this appendix an investigation where two realistic parameter scans (a.k.a. ``campaigns") are used to train ML models instead of the uniform mesh approach that is used in other parts of this paper.

As highlighted in \autoref{fig:zigzag}, there were two experimental ``campaigns" that were used to produce synthetic data -- one where thickness, focal depth and laser energy were varied assuming low pre-pulse contrast ($10^{-7}$), and another where thickness, laser energy and pre-pulse contrast were varied between $10^{-7}$ and $10^{-6}$. All the synthetic data from the two campaigns were used to train ML models. In this way, our training set includes variation in four different input parameters.

The first campaign was generated by stepping through thickness-intensity coordinates, incrementing the focal distance by \SI{3}{\micro \meter} and the thickness by \SI{0.05}{\micro \meter}, performing a full scan of focal depth values at \SI{0.5}{\micro \meter} and every integer value until \SI{5.0}{\micro \meter}.  At each point along the thickness-intensity curve, a full sweep of intensity was performed.  Since, in a real experiment, the intensity can be controlled by varying a polarizing wave plate, the synthetic data set for this investigation varied the intensity by multiplying the maximum intensity value ($10^{19}$~\unit{W . cm^{-2}}) by the cosine-squared of the wave plate angle, which was varied from $0^{\circ}$ to $70^{\circ}$ and back over the course of an intensity sweep.  The resulting sweep is depicted in \autoref{fig:zigzag}a, creating a set of 1.15 million data points.  

The second campaign was generated in a similar manner, but because, in a real experiment, neither main pulse nor pre-pulse laser intensity have an appreciable effect on target stability, both were able to be varied simultaneously.  As such, the data set was generated by incrementing thickness by \SI{0.05}{\micro \meter} from $0.5$ to $5.0$ \unit{\micro \meter}, taking a full scan of both main pulse intensity and contrast at every thickness value.  The pre-pulse contrast was varied in the same manner as the main pulse intensity, so the contrast was varied according to a cosine-squared function of another angle.  The resulting data are depicted in \autoref{fig:zigzag}b, with an overall size of 1.27 million data points.  

\begin{figure}
    \centering
    \includegraphics[width=3.25in]{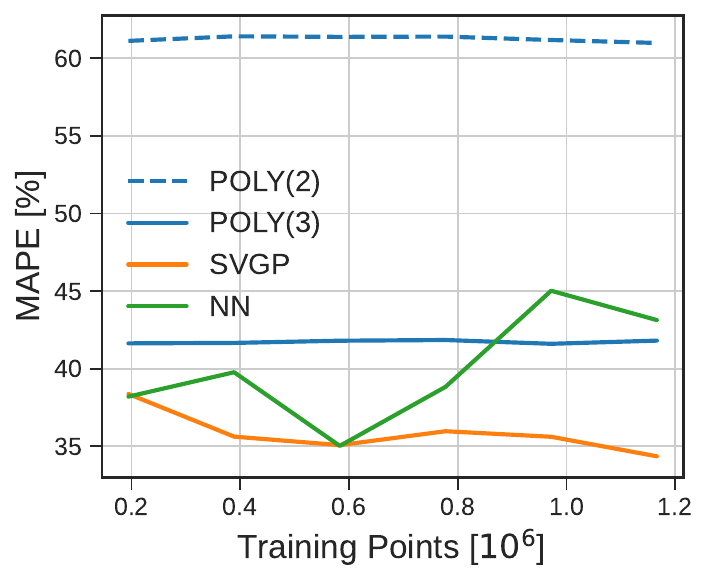}
    \vspace{-0.4cm}
    \caption{ Testing set MAPE evaluated on several ML models trained on data combined from two separate campaigns shown in \autoref{fig:zigzag}. The dashed line differs from the solid blue line in the polynomial degree. The different data splits are chosen to be approximately the same as what was shown in \autoref{fig:train_time_accuracy}.} 
    \label{fig:split_accuracy}
\end{figure}

In both campaigns, choices about how many increments to make for different parameters were influenced by a constraint that each campaign last no more than about an hour on a 1~kHz repetition rate laser system. Both campaigns assumed 10\% added gaussian noise, following the same prescription used in the body of this paper and earlier in \citet{Desai}. The combined training set, which includes data from both campaigns, has a total size of 2.42 million data points. To better compare with earlier results shown in \autoref{fig:train_time_accuracy}a in which the ML models were trained with different numbers of training points, we randomly sampled from this data set. To test the accuracy of the trained ML mdoels, we use the same testing set utilized in \autoref{fig:train_time_accuracy}a, which did not include any noise. Our results are shown in \autoref{fig:split_accuracy}. 

\autoref{fig:split_accuracy} shows that, overall, the NN and SVGP models have a much higher MAPE than was seen earlier in \autoref{fig:train_time_accuracy}a. As shown in \autoref{fig:split_accuracy}, a third order polynomial fits the data set almost as well as NN and SVGP which indicates that the NN and SVGP models are not able to fit the underlying model very well when trained on data split into two the campaigns we described. A possible improvement could be an experimental design where both the target focal position and the pre-pulse contrast are varied simultaneously, rather than keeping the contrast fixed and varying the target focal position (first campaign), and then varying the contrast while keeping the target focal position fixed (second campaign). But varying as many as four parameters simultaneously creates its own challenges for exploring a large parameter space in a relatively short amount of time ($\sim$~1-2~hours). We leave this exercise for future work.

\bibliography{ms}
\bibliographystyle{apsrev}

\end{document}